\documentclass[aps,prl,superscriptaddress,reprint,longbibliography,floatfix]{revtex4-1} 

\makeatletter
\newif\ifpreprintoption
\@ifclasswith{revtex4-1}{preprint}{\preprintoptiontrue}{\preprintoptionfalse}
\makeatother
\newcommand{\ifpreprint}[2]{\ifpreprintoption #1\else #2\fi}

\usepackage{tabularx}
\usepackage{multirow}
\usepackage{amsmath,amssymb}
\usepackage{lmodern}
\usepackage{graphicx}
\usepackage{epstopdf}
\usepackage[colorlinks=true,linkcolor=red,bookmarks=true,citecolor=blue,urlcolor=blue]{hyperref}
\usepackage[T1]{fontenc}
\usepackage{color}
\usepackage[separate-uncertainty=true]{siunitx}
\usepackage{lipsum}
\usepackage[plain]{fancyref}
\usepackage{soul}

\begin{document}
\title{Generation and Characterization of Attosecond Micro-Bunched Electron Pulse Trains via Dielectric Laser Acceleration}

\author{Norbert Sch\" onenberger}
\email[]{norbert.schoenenberger@fau.de}

\author{Anna Mittelbach}
\author{Peyman Yousefi}
\author{Joshua McNeur}
\affiliation{Department of Physics, Friedrich-Alexander Universit\" at Erlangen-N\" urnberg (FAU), \\ Staudtstra\ss e 1, 91058 Erlangen, Germany}

\author{Uwe Niedermayer}
\affiliation{Technische Universit\" at Darmstadt, Institut f\" ur Teilchenbeschleunigung und Elektromagnetische Felder (TEMF) \\ Schlossgartenstra\ss e 8, 64289 Darmstadt, Germany}

\author{Peter Hommelhoff}
\email[]{peter.hommelhoff@fau.de}

\affiliation{Department of Physics, Friedrich-Alexander Universit\" at Erlangen-N\" urnberg (FAU), \\ Staudtstra\ss e 1, 91058 Erlangen, Germany}

\date{\today}

\begin{abstract}
Dielectric laser acceleration is a versatile scheme to accelerate and control electrons with the help of femtosecond laser pulses in nanophotonic structures. We demonstrate here the generation of a train of electron pulses with individual pulse durations as short as $270\pm80$ attoseconds(FWHM), measured in an indirect fashion, based on two subsequent dielectric laser interaction regions connected by a free-space electron drift section, all on a single photonic chip. In the first interaction region (the modulator), an energy modulation is imprinted on the electron pulse. During free propagation, this energy modulation evolves into a charge density modulation, which we probe in the second interaction region (the analyzer). These results will lead to new ways of probing ultrafast dynamics in matter and are essential for future laser-based particle accelerators on a photonic chip. 
\end{abstract}


\maketitle

Ultrashort electron pulses find various applications in research and technology, including ultrafast diffraction~\cite{Siwick2003,Baum2007b,Morimoto2018}, ultrafast electron microscopy~\cite{Zewail2009,Schliep2017,RubianoDaSilva2018,Berruto2018}, as well as ultrafast photon generation~\cite{Ackermann2013}. Many of these techniques operate with electron pulse durations in the realm of femtoseconds.

In order to resolve processes taking place on atomic time scales in atoms or molecules or on electronic time scales in solids, electron pulses with attosecond duration are highly sought after. The temporal resolution of laser-triggered electron sources is usually limited by the temporal duration of the electron-releasing laser pulses and subsequent dispersive broadening of the electron pulses. Typical electron pulse durations at the sample are in the range of \SI{30}{\femto\second} to \SI{1}{\pico\second}~\cite{Siwick2003,Baum2007b,Morimoto2018,Zewail2009,Schliep2017,RubianoDaSilva2018,Berruto2018,Ackermann2013,Kozak2018SEM}. Schemes have been proposed and demonstrated to compress the electron pulses at the sample, see for example~\cite{Kealhofer2016,Sciaini2011}. The shortest pulse duration demonstrated this way is \SI{6}{\femto\second} so far~\cite{Zhao2018}. 

The temporal resolution can be increased significantly by utilizing directly the optical carrier field of ultrashort laser pulses. Energy modulation of the free electrons via optical fields, for example, can be accomplished in several different schemes, leading to electron pulse trains with sub-optical cycle bunchlet duration.  One such method is to utilize ponderomotive forces~\cite{Baum2007,Hilbert2009a}. Recently, microbunches as short as 260\,as have been realized this way~\cite{Kozak2018,Kozak2018a}.  In another scheme, the inverse free electron laser (IFEL) process has been used, where microbunch durations as short as 410\,as have been demonstrated~\cite{Sears2008a}. Finally, optical nearfields can be used to transfer momentum from a light field to free electrons. With nearfields generated by (metallic) plasmonic nanostructures, pulse durations as short as 655\,as  have been reached~\cite{Priebe2017}.

\begin{figure*}[!ht]
\centering
\includegraphics[width=0.8\textwidth]{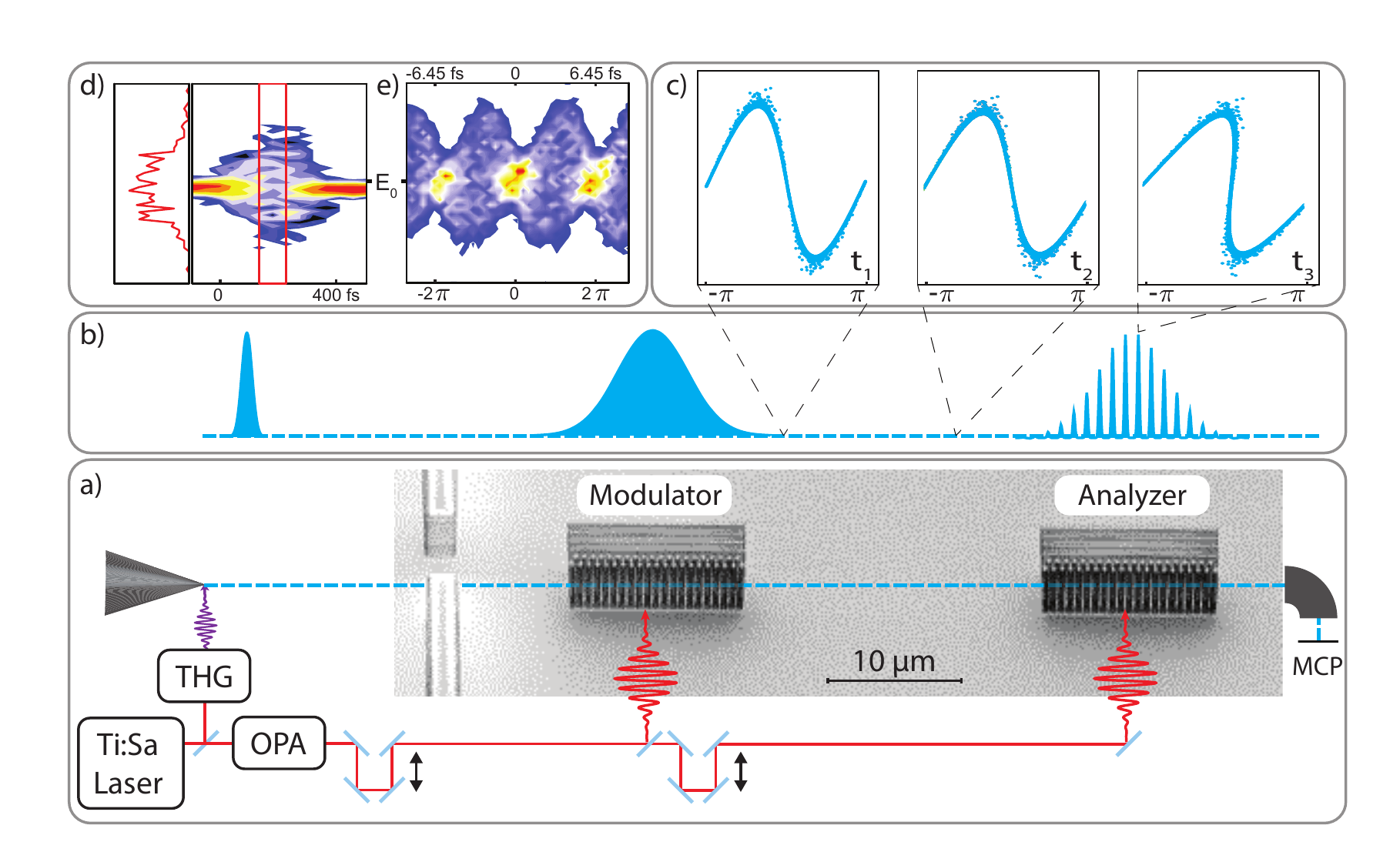}
\caption{\label{fig:setup} \ifpreprint{ \footnotesize}{} Sketch of the experimental setup with modulator and analyzer structure and sketches of the electron phase space behavior. a) Laser-emitted electrons are focused into the center of the channel of the first dielectric laser acceleration structure, comprised of two rows of pillars, the modulator. An SEM image of modulator and analyzer structure can be seen in the background of this sketch. After the electrons have propagated through the analyzer structure, their energy is measured with a magnetic deflection spectrometer. (b) Sketch of the evolution of the electron pulse duration. At the source, the electron pulse duration resembles that of the triggering UV laser pulse ($\sim \SI{100}{\femto\second}$). During propagation through the electron column, trajectory effects increase the electron pulse duration to roughly \SI{400}{\femto\second} at the modulator. The pulsed laser beam acting on each arriving electron pulse modulates the energy of the electrons. During subsequent propagation, the energy modulation leads to a density modulation. At the temporal focus, the minimum electron pulse duration of each bunchlet is reached. The position of the temporal focus depends on the amplitude of the energy modulation in the modulator. Here microbunching at the position of the analyzer is shown. (c) Sketch of the phase space evolution during the electron drift. The vertical axis denotes the energy of the electrons plotted over one cycle ($- \pi \dots \pi \equiv 6.45\,$fs). The faster higher energy electrons catch up with the slower electrons, forming the micro-bunched pulse train.  (d) Example spectrogram  of the electrons after interaction in the modulator only (laser intensity of \SI{3e11}{\watt\per\centi\meter\squared}). The red curve shows the homogeneous broadening inside the red region. (e) Example spectrogram with modulator and analyzer structure illuminated (\SI{1.5e10}{\watt\per\centi\meter\squared} in the modulator, \SI{2.5e10}{\watt\per\centi\meter\squared} in the analyzer). The periodicity with the optical period of $6.45$\,fs and sub-optical cycle duration features are clearly visible.}
\end{figure*}

We here use dielectric (transparent) nano-photonic structures made from silicon. They are extremely versatile and easy to produce, even in large numbers. These structures are utilized to generate an optical nearfield allowing efficient momentum transfer from the lightfield into the electron beam over a prolonged interaction distance, which other schemes cannot provide. Because these structures only vary the phase of the optical field on sub-optical cycle dimensions, the interaction of the light field with the structure can be modelled as a purely dispersive effect. Hence, light absorption hardly takes place in these structures, allowing us to reach high laser damage thresholds in excess of \SI{2}{\giga\volt\per\meter}, corresponding to peak intensities of \SI{5e11}{\watt\per\centi\meter\squared}. In addition to this high damage threshold, these structures are highly advantageous over other schemes and structures because of their broad functionality that can be encoded into the nanostructure.

Various dielectric structures for laser-driven particle acceleration have been proposed (\cite{England2014, Wootton2017a}, and references therein). In 2013, dielectric laser acceleration was shown experimentally, demonstrating phase-synchronous acceleration of charged particles with light fields~\cite{Peralta2013, Breuer2013}. Quickly thereafter, various other functionalities have been realized based on this scheme of phase-synchronous interaction of nearfields generated in dielectric structures and fast electron pulses, both at relativistic and non-relativistic energies. Examples include the deflection, focusing and streaking of an electron beam~\cite{Wootton2017,Kozak2017,McNeur2018,Black2019}. With all these individual building blocks available, and with the demonstration of two concatenated structures~\cite{Kozak2017}, the concept of a particle accelerator on a photonic chip is now within reach. Importantly, acceleration of electrons in infinitely long structures with negligible electron loss has recently been demonstrated numerically based on alternating phase focusing \cite{niedermayer2018}. 
In this letter, we show that by carefully controlling the phase space dynamics of a pulsed electron beam, sub-optical cycle bunching and attosecond bunch generation can be achieved \cite{Niedermayer2017}. For this, we imprint an energy modulation periodic with the driving optical period of \SI{6.45}{\femto\second} on each 400\,fs long electron pulse in a first nearfield interaction section called the modulator (Fig.~\ref{fig:setup} a). 
This nearfield can be described by the following formula \cite{Breuer2014a}:
\begin{widetext}
\begin{equation}
\label{field}
    E(x,z)=c\left(\begin{array}{cc}
     -\frac{1}{\beta\gamma^2}\:(\:C_s\: \mathrm{sinh}\,(k_xx)\: + \:C_c\: \mathrm{cosh}\,(k_xx))\:\mathrm{cos}\,(k_zz-\omega t)\\
        0  \\
       \frac{1}{\beta\gamma}\:(\:C_s\: \mathrm{cosh}\,(k_xx)\: + \:C_c\: \mathrm{sinh}\,(k_xx))\:\mathrm{sin}\,(k_zz-\omega t)\\
    \end{array}\right)
\end{equation}
\end{widetext}
where $C_c = 0$ and $C_s$ is proportional to the field amplitude. $x$ is the transverse coordinate and $z$ is the electron propagation direction.

After this, the energy-modulated electron pulse propagates freely to the second nearfield interaction section called the analyzer (structure and field is identical to the modulator field shape in Eq.\eqref{field}). During this drift, the energy modulation develops into a density modulation, probed in the analyzer section and diagnosed with the a dipole magnet electron spectrometer. We obtain feature-rich electron spectrograms, which show the electron energy versus the time delay (up to an offset) between modulator and analyzer laser pulses. By comparing these spectrograms to numerically obtained ones, we can clearly show the sub-optical cycle, attosecond electron pulse duration. 

The experiments are performed in an ultrafast scanning electron microscope (USEM). Laser pulses from an amplified titanium:sapphire laser with pulse durations of \SI{100}{\femto\second} and a repetition rate of \SI{1}{\kilo\hertz} are fed into an optical parametric amplifier (OPA). A part of the fundamental output is used to generate the third harmonic to photoemit electrons from the Schottky-type emitter in a modified commercial electron microscope, which serves as the electron source. The dielectric structures are illuminated by laser pulses generated in the OPA with a wavelength of \SI{1932}{\nano\meter} and a pulse duration of \SI{650}{\femto\second}, obtained via a Fabry Perot filter. The relative phases or time differences of the pulsed laser beams impinging on the modulator and analyzer structure as well as on the electron source are precisely adjusted via delay stages. The electron microscope is operated at $\beta = \frac{v_e}{c} = 0.32$ corresponding to an energy of \SI{28.4}{\kilo\electronvolt}. The spot size is approx. \SI{50}{\nano\meter}, with a divergence angle of approx. \SIrange{1}{2}{\milli\radian}. The focus of the electron beam is adjusted to be as close as possible to the center of the structure. Since the electron pulses experience temporal broadening due mainly to trajectory effects inside of the electron column, the pulses have a duration of $\sim$\SI{~400}{\femto\second} when they reach the dielectric structure \cite{Kozak2018SEM}.  It is mainly because of the small laser repetition rate and the use of a commercial SEM as electron source that the electron count rate is only 1-10 electrons per second here, implying that one 400\,fs long electron pulse contains less than one electron on average. This is because the SEM is optimized for high resolution and image quality, so the (usually DC) electron beam is heavily filtered by various apertures. More details of the setup can be found in~\cite{Kozak2018SEM}.

\begin{figure*}[!ht]
\centering
\includegraphics[width=0.95\textwidth]{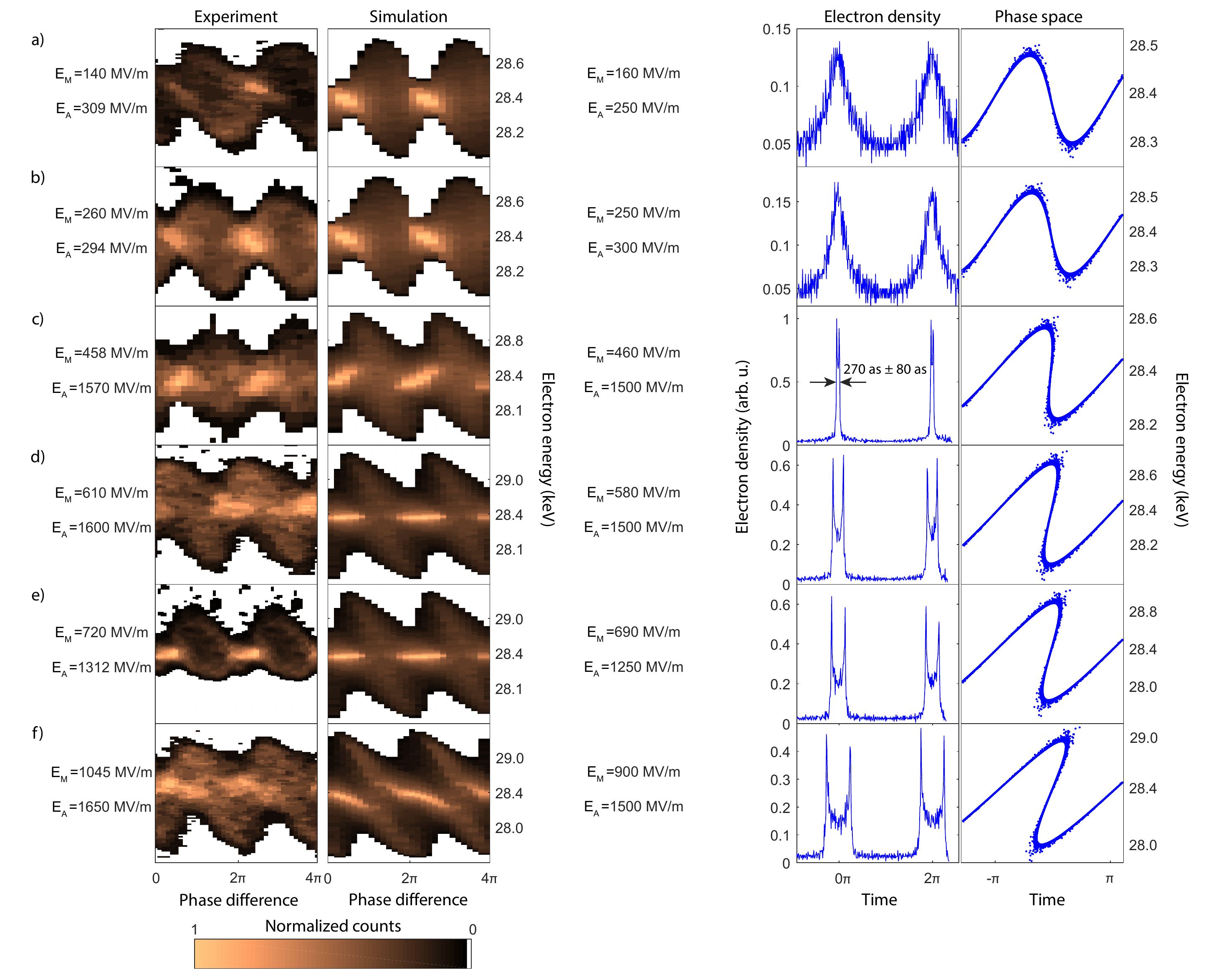}
\caption{\label{fig:results}  \ifpreprint{ \footnotesize}{} Spectrograms, electron density and phase space at the analyzer structure. The left column shows experimental data, next to it we show simulated spectrograms and the third and forth column show the electron density distribution and phase space at the position of the analyzer, respectively. The color scale of the first two columns shows electron counts, normalized to the maximum count of electrons for each data set individually. Peak fields are given next to the spectra. Under-bunching is clearly visible in the first two rows, where the temporal focus lies behind the analyzer. Shortest micropulses are shown in c), with a pulse duration of $\SI{270+-80}{\atto\second}$. Various degrees of over bunching are displayed in the last three rows. The forth column shows the phase space distribution of the microbunches: the vertical axes represent the electrons' energy. When the maximum and minimum of the modulation coincide in time, the temporal focus is reached. After that the characteristic over-bunching shape is formed, when the high energy electrons have passed the slow electrons in the analyzer and the temporal focus lies in front of it.}
\end{figure*}

We chose dual pillar structures etched into silicon as building blocks of the dielectric structures for their ease of manufacturing and laser in-coupling~\cite{Leedle2015}. Modulator and analyzer structure are identical and \SI{13.2}{\micro\meter} long. The distance from the end of the modulator to the center of the analyzer constitutes the drift section, which was chosen to be \SI{30}{\micro\meter}. More details about the structure are included in the supplementary material. After the analyzer structure, the electrons propagate into a magnetic deflection spectrometer with an energy resolution of \SI{\sim 40}{\electronvolt}~\cite{Kozak2018SEM}. This way, we record spectrograms by plotting electron spectra versus the time delay between the pulsed laser beams impinging on modulator and analyzer structure (Fig.\ref{fig:setup}). The pulsed laser beams are focused on the structures down to a spot size of roughly \SI{14\pm 0.5}{\micro\meter} ($1/e^2$ intensity radius). With average laser powers of \SIrange{30}{100}{\micro\watt}, we generate peak intensities of \SIrange{1.5e10}{4.7e10}{\watt\per\centi\meter\squared}, corresponding to peak optical fields of \SIrange{340}{600}{\mega\volt\per\meter} on the modulator. With a structure factor of $\sim 0.1$~\cite{Yousefi2019}, which determines the conversion of incident field to the synchronous mode, the resulting peak acceleration gradient acting on the electrons is \SIrange{34}{60}{\mega\electronvolt\per\meter}. Damage usually sets in at a field strength of around \SI{2}{\giga\volt\per\meter}, corresponding to peak intensities of \SI{5e11}{\watt\per\centi\meter\squared}. The structure has been designed for an intermediate gradient of \SIrange{20}{25}{\mega\electronvolt\per\meter} corresponding to a longitudinal focal length of \SI{30}{\micro\meter}. This is a compromise between defocusing, energy spread and the ability to properly separate the two laser spots. The longitudinal focal length,
\begin{equation}
    L = \frac{\lambda_g}{2\pi}\cdot \beta^2\gamma^3 \cdot mc^2 \cdot \frac{1}{\Delta E}
\end{equation}
with $\lambda_g$ the periodicity of the structure, $\beta$ the speed of the electrons in units of the speed of light and $\gamma$ the Lorentz factor, is defined as the distance from the end of the first structure to the plane where of shortest micropulse duration, i.e. the plane in which the imprinted velocity modulation has evolved until the fast electrons have caught up with the slow ones \cite{Niedermayer2017}.


Fig.~\ref{fig:results} shows experimental data, simulation results, retrieved time traces and phase space diagrams for various laser intensities in the modulator. The measured spectrograms include effects from both the modulator and analyzer, rendering a direct extraction of the electron time structure at the analyzer position difficult. For this reason, we compare the measured spectrograms with numerically obtained ones to indirectly measure the micropulse length. The numerical spectrograms are based on a finite difference time domain (FDTD) simulation of the optical nearfields inside of the nanophotonic structures~\cite{lumerical}, while the electron tracking was performed with a Runge-Kutta motion solver~\cite{gpt}. From the simulation results, we also obtain the electron density and the phase space distribution shown in the two rightmost columns in Fig.~\ref{fig:results}. 

From top to bottom, the laser intensity in the modulator structure increases. In the first two cases, Fig.~\ref{fig:results} a) and b), the laser intensity in the modulator is below the intensity needed to produce fully bunched electrons at the analyzer structure. Hence, the temporal focus of the micro-bunches lies after the center of the analyzer. For the parameters shown in Fig.~\ref{fig:results} c), the laser intensity is almost ideally matched to the drift length so that the electron pulses are close to the minimal pulse duration. The onset of over-bunching is, however, already discernible. When we increase the laser intensity in the modulator even further [Fig.~\ref{fig:results} d) - f)], the temporal focus shifts closer to the modulator, resulting in clear over-bunching at the position of the analyzer. 

To find the best matching numerical spectrograms, we have simulated various parameter sets close to the experimental ones. This process and the concomitant pulse duration extraction are detailed in the supplemental material. 
Importantly, the length of the electron micro-bunch duration can not be directly inferred from features contained in the recorded spectrograms. This is because the relatively high energy spread induced in the modulator structure causes significant bunch evolution even in the analyzer structure since the temporal depth of focus is so narrow. A unique best matching solution for the spectrogram in Fig.~\ref{fig:results} c) is found for these parameters: E$_M$=\SI{460}{\mega\volt\per\meter}, the incident field strength on the modulator structure, and E$_A$=\SI{1500}{\mega\volt\per\meter}, the incident field on the analyzer. The resulting micro-bunch duration (full width at half maximum, FWHM) is $\SI{270+-80}{\atto\second}$. The error is derived from the comparison with adjacent simulation results. In the vicinity of our shortest measured micro-bunches, a variation of \SI{\pm20}{\mega\volt\per\meter} in the modulator results in a change of \SI{\pm80}{\atto\second} in the analyzer structure. Hence, we conservatively estimate the measurement error to be \SI{80}{\atto\second}. An extensive sweep, which was performed with step sizes of \SIrange{40}{300}{\mega\volt\per\meter} for the modulator fields and \SI{250}{\mega\volt\per\meter} for the analyzer can be found in the supplementary materials. Within this grid of simulations we can uniquely identify the simulated spectrum that matches the experiment best.

\begin{figure}[!ht]
\centering
\includegraphics[width=0.4\textwidth]{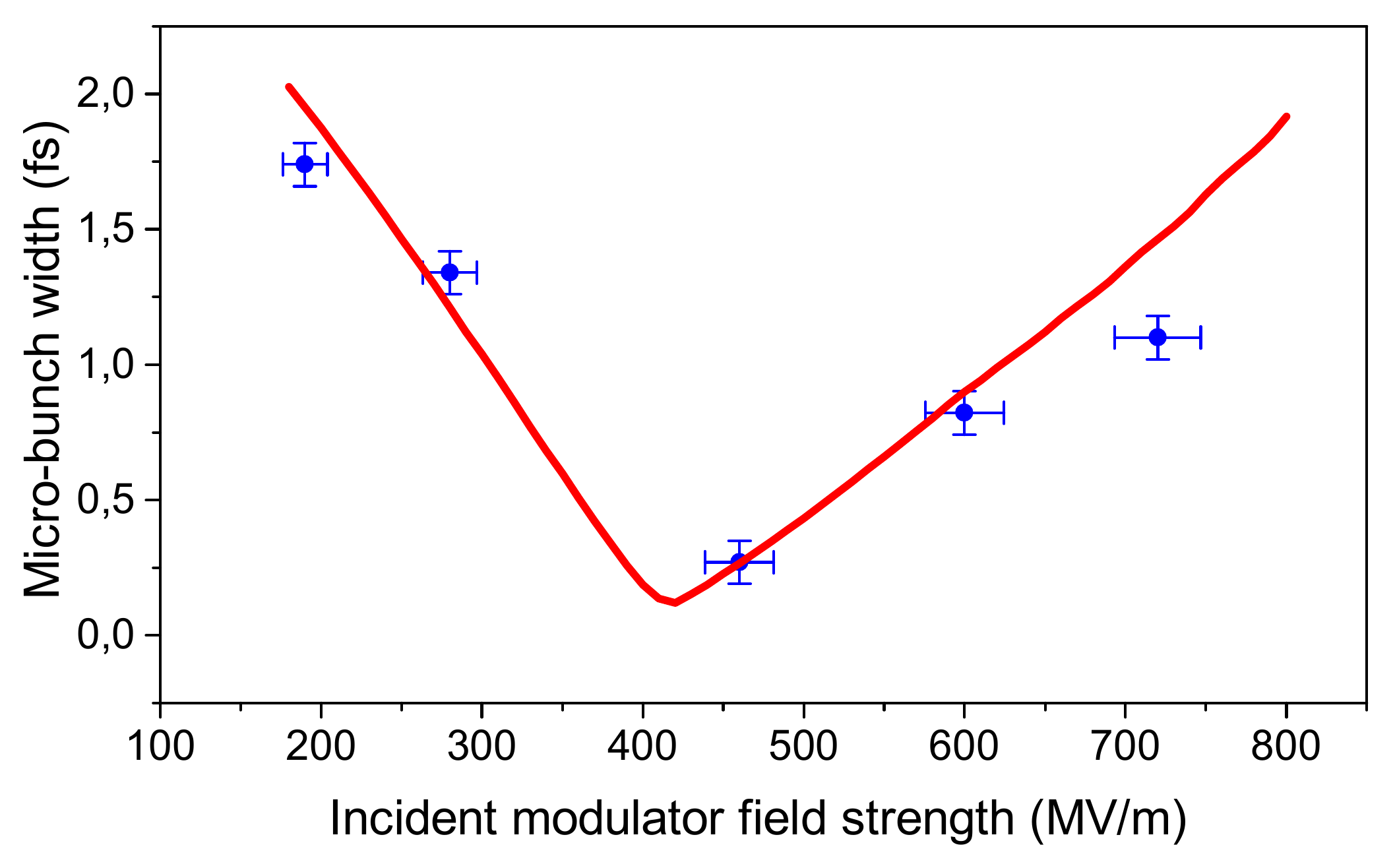}
\caption{\label{fig:linet} Microbunch duration versus incident field strength on the modulator after a fixed drift length of \SI{30}{\micro\meter}. The blue datapoints show the measurements with their respective micro-bunch lengths, while the red curve shows the simulation results, with a minimum of \SI{125}{\atto\second}.}
\end{figure}

The experimentally achieved micro-bunch duration in the almost ideally bunched case of \SI{270 +- 80}{\atto\second} corresponds to just $4\,\%$ or 270\,mrad of the optical driving period. In the simulation, we find that a minimal electron micro-bunch duration of \SI{125}{\atto\second} should be feasible with the scheme and laser pulse parameters employed here (Fig.~\ref{fig:linet}). With incident fields close to the damage threshold and shorter drift spaces, one might even consider reaching the single digit attosecond range. Simulations indicate that the structures used here would produce \SI{7}{\atto\second} (FWHM) micro-bunches with an incident field on the modulator of \SI{1.5}{\giga\volt\per\meter} after a drift space of \SI{2}{\micro\meter}, when the initial energy spread is small (< \SI{1}{eV}). Note that spectra approaching a double hump structure resulting from a sinusoidal modulation have been observed \cite{Yousefi2019}. This indicated that we can observe the required beam dynamics.

To summarize, we have demonstrated attosecond micro-bunch train generation with individual bunchlet durations as short as \SI{270 +- 80}{\atto\second}. Simulations show that the shortest  micro-bunch duration with the current nanostructure could reach \SI{125}{\atto\second}. Even shorter bunches can be achieved by reducing the drift section and using higher field strengths in the modulator section.
The resulting micro-bunch trains could already be utilized to probe coherently pumped processes in a stroboscopic fashion. An increase in the available currents by orders of magnitude is straightforward by going to higher repetition rate laser sources (commercially available) and to better matched mini- or even micro-electron optics. Furthermore, advances in the fields of optical field-driven particle accelerators require sub-optical cycle-bunched electrons to be injected into the proper phase space region for the acceleration to be efficient and lossless. The precisely defined injection phase demonstrated here paves the way to matched injection into the acceptance of a scaleable DLA using technologies like the aforementioned alternating phase focusing. This enables to not only modulate electron energy, i.e. to have a beam with a big energy spread, comprised of accelerated and decelerated electrons, but to produce a net accelerated beam, where a substantial portion of the electrons is shifted to a higher energy with an energy spread significantly lower than that of a purely modulated beam. Our work will hence enable both new time-resolved electron-based imaging as well as building new and efficient optical particle accelerators.  Similar results are reported in~\cite{Black2019a}. 

\begin{acknowledgments}
This work was funded by the Gordon and Betty Moore Foundation (GBMF) through Grant No. GBMF4744  ''Accelerator on a Chip International Program-ACHIP'', and BMBF via 05K16WEC and 05K16RDB.
\end{acknowledgments}

\end{document}